\documentclass[twocolumn,times,tighten]{aastex63}
\received{---}
\revised{----}
\accepted{--- }          

\usepackage{color}
\usepackage{graphicx}
\usepackage{amssymb}
\usepackage{array}
\usepackage{color}
\usepackage{subfigure} \usepackage{threeparttable}
\usepackage{natbib}
\bibpunct{(}{)}{;}{a}{,}{,}

\usepackage{enumitem}
\usepackage{hyperref}
\usepackage{verbatim}
\usepackage{amsmath,amstext,mathrsfs}
\usepackage[all]{hypcap} 
\usepackage{afterpage}    
\usepackage{marginnote}
\usepackage{makecell}
\usepackage{subfigure}
\usepackage{xfrac}
\usepackage{ifxetex,ifluatex}
\usepackage{multirow}
\usepackage{lineno}

\newcommand*{\kh}{} 

\def \sun {$_{\odot}$}

\shorttitle{Applying the VGT in NGC~1333}
\shortauthors{Soam et al.}

\begin{document}

\title{Applying the Velocity Gradient Technique in NGC~1333: Comparison with Dust Polarization Observations}

\correspondingauthor{Archana Soam}
\email{archana.soam@iiap.res.in}

\author[0000-0002-6386-2906]{Archana Soam}
\affiliation{Indian Institute of Astrophysics, II Block, Koramangala, Bengaluru 560034, India}

\author[0000-0003-1683-9153]{Ka Ho Yuen}
\affiliation{Theoretical Division, Los Alamos National Laboratory, Los Alamos, NM 87545, USA}
\email{kyuen@lanl.gov}

\author{Ian Stephens}
\affiliation{Department of Earth, Environment, and Physics, Worcester State University, Worcester, MA 01602, USA}

\author{Chi Yan Law}
\affiliation{Department of Earth and Space Sciences, Chalmers University of Technology, Sweden}
\affiliation{European Southern Observatory, Karl-Schwarzschild-Strasse 2, D-85748 Garching, Germany}

\author{Ka Wai Ho}
\affiliation{Theoretical Division, Los Alamos National Laboratory, Los Alamos, NM 87545, USA}
\affiliation{Department of Astronomy, The University of Wisconsin-Madison, Wisconsin, WI 53706, USA}

\author[0000-0002-0859-0805]{Simon Coud\'e}
\affiliation{Department of Earth, Environment, and Physics, Worcester State University, Worcester, MA 01602, USA}
\affiliation{Center for Astrophysics $\vert$ Harvard \& Smithsonian, 60 Garden Street, Cambridge, MA 02138, USA}

\begin{abstract}
Magnetic fields (B-fields) are ubiquitous in the interstellar medium (ISM), and they play an essential role in the formation of molecular clouds and subsequent star formation. However, B-fields in interstellar environments remain challenging to measure, and their properties typically need to be inferred from dust polarization observations over multiple physical scales. In this work, we seek to use a recently proposed approach called the Velocity Gradient Technique (VGT) to study B-fields in star-forming regions and compare the results with dust polarization observations in different wavelengths. The VGT is based on the anisotropic properties of eddies in magnetized turbulence to derive B-field properties in the ISM. We investigate that this technique is synergistic with dust polarimetry when applied to a turbulent diffused medium for the purpose of measuring its magnetization. Specifically, we use the VGT on molecular line data toward the NGC~1333 star-forming region ($\rm ^{12}CO$, $\rm ^{13}CO$, $\rm C^{18}O$, and $\rm N_{2}H^{+}$), and we compare the derived B-field properties with those inferred from 214 and 850~$\mu$m dust polarization observations of the region using SOFIA/HAWC+ and JCMT/POL-2, respectively. We estimate both the inclination angle and the 3D Alfvénic Mach Number $M_A$ from the molecular line gradients. Crucially, testing this technique on gravitationally bound, dynamic, and turbulent regions, and comparing the results with those obtained from polarization observations at different wavelength, such as the plane-of-the-sky field orientation, is an important test on the applicability of the VGT in various density regimes of the ISM. 
\end{abstract}




\section{Introduction} \label{sec:intro}

Magnetic fields (B-fields) have important roles in various astrophysical phenomena, such as regulating the evolution of large interstellar filamentary structures and their embedded cores \citep{2016A&A...592A..54A}, and the rate of star formation \citep[e.g.][]{2022MNRAS.514.3024L}. The gravitational stability of these cores, and thus their likelihood to host star formation, can be measured by comparing their magnetic, gravitational, and kinetic energy \citep[see][]{federrath2015}. Recent magnetohydrodynamic (MHD) simulations \citep{ps2018} suggest a complex evolutionary process with dense velocity coherent fibers collapsing into chains of cores, which resembles observations of molecular clouds  \citep[e.g., L1495;][]{marsh2016}. 

The decades-long accepted procedure of mapping B-fields using polarized dust emission has proven to be incredibly useful and productive over the years. The alignment of grains in the interstellar medium (ISM) has been found to result from a process called Radiative Alignment Torques \citep[e.g.,][]{1976Ap&SS..43..291D,2007MNRAS.378..910L,2015ARA&A..53..501A}. This model predicts that the observed polarization is due to asymmetric dust grains made to rotate due to radiative torques, thus aligning their shortest axes parallel to the ambient B-field. The polarized emission from these dust grains are therefore perpendicular to the B-field lines. Due to high dust extinction (optical/NIR) and low resolution (e.g. 5$\arcmin$ with $Planck$), previous work on interstellar B-fields has been mostly limited to diffuse and/or nearby clouds, and only recently have additional studies started systematically probing denser filamentary infrared dark clouds (IRDCs) using (sub)millimeter polarization observations \citep[e.g.,][]{pillai2015, Liu2018, soam2019, stephen2022}. The orientation of B-fields relative to interstellar density structures has been studied in different context, such as at galactic scales with $Planck$ \citep{soler2013}, or in selected Gould Belt molecular clouds \citep{hbli2013}. Theoretically, we expect lower-density structures that are not subject to strong gravitational collapse to be generally parallel to B-fields \citep{YL17a,soler2017}. While dust polarization has been successful in mapping B-field structures in IRDCs, it would still be beneficial to have independent techniques available to confirm their derived magnetic properties. 

The Velocity Gradient Technique \citep[VGT, see][]{GL17, YL17a, YL17b, LY18a, lazarian2018b} is a recently proposed method for estimating both the {\it plane-of-sky} B-field direction and strength in interstellar turbulent media. The technique is based on the fact that MHD turbulence tends to create anisotropic eddies along the local B-field directions \citep{GS95}. In the classical sense, fluid motions perpendicular to B-field lines will tend to twist and tangle up the B-field lines without constant energy input. The concept of stochastic turbulent diffusion \citep{LV99} allows fluid parcels with a pinched  B-field to rotate perpendicular to the mean B-field direction without the field lines to entangle.
Therefore, a high-resolution position-position-velocity (PPV) cube of spectral line observations can be used to estimate the B-field orientation in a region \citep{cattail}. The VGT has already been used to study B-field orientations in comparison with polarization observations in diffuse gas (e.g. HI: \citealt{YL17a}, CO : \citealt{LY18a}, Smith cloud, \citealt{hu2019}). Studies show that the B-field directions as determined by the VGT method and $Planck$ data have a strong correlation \citep{yue2019Nat}. However, the velocity and B-field structures in high-density regions may be complicated by other factors, such as infall and outflows. Therefore, a question persists on the applicability of VGT on smaller-scale, denser regions. Can we predict the orientation of B-fields on these scales using PPV cubes, and will these orientation match with the ones mapped from polarization observations? Are velocity gradients aligned with B-field orientation in a gravitationally collapsing region, which is opposite to what is seen in diffuse elongated structures?
 
\begin{figure*}
\centering
\includegraphics[width=0.495\textwidth]{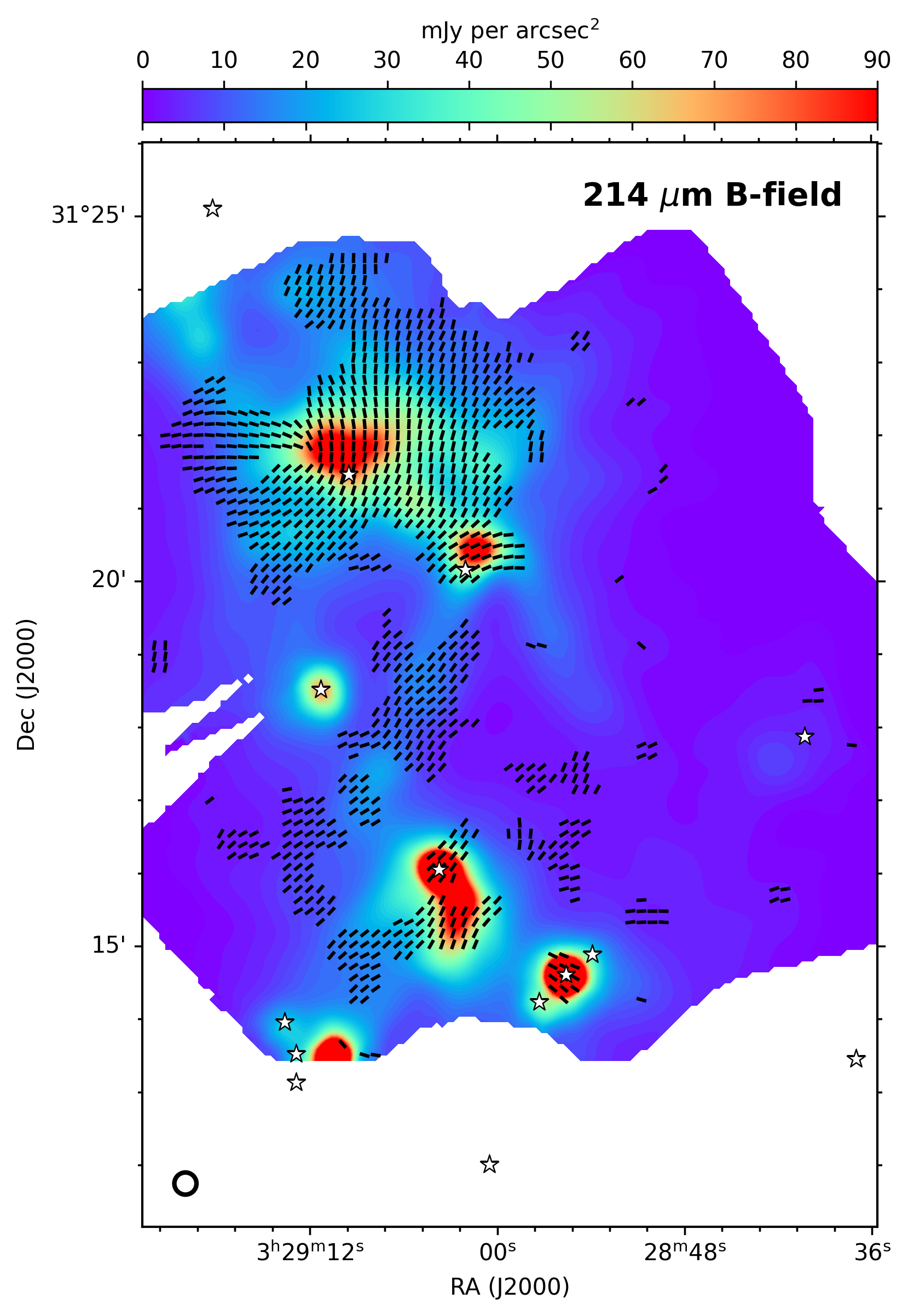}
\includegraphics[width=0.495\textwidth]{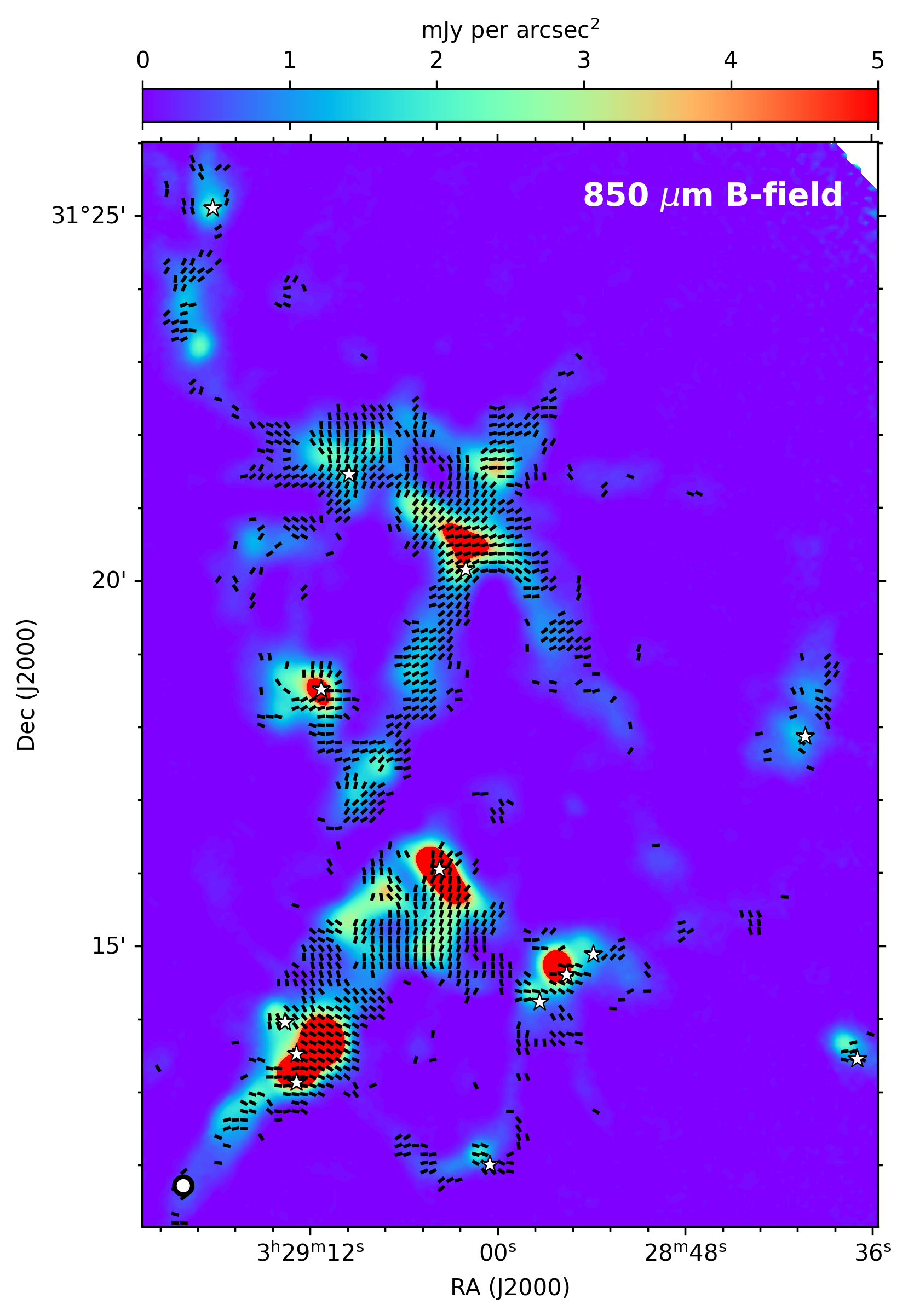}
\caption{{\bf Left panel}: The B-field morphology traced by the SOFIA/HAWC+ $214~\mu m$ dust polarization observations of NGC\,1333 plotted on the Stokes~I total intensity map. The plotted line-segments correspond to $P/\sigma_P > 3$, and only one in two vector is plotted for clarity.  {\bf Right panel}: B-field map obtained from $850~\mu m$ polarizations observations plotted on the Stokes~I total intensity, as adapted from \citet{doi2020}. Young Stellar Objects identified from \citet{sandell2001} are shown in both panels. The SOFIA/HAWC+ and the JCMT/POL-2 beam sizes are shown in the lower left corner of their respective panel.}
\label{fig:lic_plot}
\end{figure*}

To address these questions, we chose to analyze the filamentary region NGC~1333 in the Perseus molecular cloud, which is located $\sim$300\,pc from the sun \citep{Zucker2018}. The magnetic properties in NGC~1333 were previously studied with dust polarization measurements from the B-Fields In Star-forming Region Observations (BISTRO) survey using the James Clerk Maxwell Telescope (JCMT) \citep{doi2020}. In this work, we complement the available published BISTRO observations in the submillimeter (850\,$\mu$m) with our far-infrared (214\,$\mu$m) dust polarization measurements using the Stratospheric Observatory for Infrared Astronomy (SOFIA). Figure~\ref{fig:lic_plot} shows the derived plane-of-the-sky B-field maps for each data set. We also use molecular lines observations such as $\rm ^{12}CO$, $\rm ^{13}CO$, $\rm C^{18}O$, and $\rm N_{2}H^{+}$ from the JCMT and CARMA archives to compute velocity gradients. 

This paper is organized as follows: in Section~\ref{sec:obs}, we describe the polarimetric and spectroscopic observations; in Section~\ref{sec:analysis}, we present the analysis using the Velocity Gradient Technique (VGT) and gradient statistics; in Sections ~\ref{sec:result} and \ref{sec:malik}, we present our results and calculations of various parameters; in Section~\ref{sec:discussion}, we discuss the significance of our findings, and finally in Section~\ref{sec:conclusion}, we conclude our work.

\section{Data acquisition and reduction} \label{sec:obs}

{\kh 
\begin{deluxetable*}
{ccccccc}
\tablecaption{Summary of the observed data used in this work. \label{tab:result_obs}}
\tablehead{
Telescope &
Species & Transition&
$\lambda$/$\nu$
& & FWHM$_{\rm beam}$
& Ref}
\startdata 
&& &$\mu m$/GHz& & $(^{\prime\prime})$& \\
\hline\hline 
&& & Molecular emissions & & &\\
\hline
JCMT/HARP & \makecell{$\rm {}^{12}CO$ \\ $^{13}$CO \\ $C^{18}$O}& J=3--2 & \makecell{$345.796\,$GHz \\ $330.587\,$GHz  \\ $329.331 \,$GHz}&& $\rm 14 \farcs 2$ &(1)\\
CARMA & N$_2$H$^+$ &  J=1--0 & 93.174 &  & $\rm 7\farcs2\,\times\,5\farcs4$ &(2)\\
\hline
&& & Dust polarization & & &\\
\hline
SOFIA/HAWC+ &  &  & 214  & & 18$\farcs$2 & (3)\\
JCMT/POL-2 &  &  &850  & & 14$\farcs$2 &  (4)\\
\hline 
\hline\hline
\enddata
\begin{tablenotes}[flushleft]
  \item[](1). Archival data ID: M06BGT02 (PI: Richard, Hills)
  \item[](2). CLASSY: \citet{2014ApJ...794..165S}
  \item[](3). SOFIA proposal ID: $06\_0098$ (PI: Stephen, Ian).
  \item[](4). \citet{doi2020}.
\end{tablenotes}
\end{deluxetable*}
}

\subsection{Dust polarization at 214 $\mu$m with HAWC+} \label{sec:optPol}

We observed NGC~1333 with the High-Resolution Airborne Wide-band Camera Plus (HAWC+) of the Stratospheric Observatory For Infrared Astronomy (SOFIA) on flights 507 (2018/09/20) and 511 (2018/09/27) under SOFIA program 06\_0098 (PI: Ian Stephens). SOFIA was an airborne observatory equipped with a 2.5~m primary mirror and jointly operated by the National Aeronautics and Space Administration (NASA) and the German Aerospace Center (DLR) from 2010 to 2022. HAWC+ is a far-infrared polarimetric camera capable of measuring the polarized thermal emission of interstellar dust at 54, 89 154, and 214~$\mu$m \citep{Harper2018}. Specifically, we targeted four fields in NGC~1333 with HAWC+ in Band~E polarization (214~$\mu$m) while using a Chop-Nod imaging mode. This mode uses the secondary mirror of SOFIA's telescope array to alternate, or ``chop'', at a frequency of $10.2$~Hz between an ``On'' and an ``Off'' position separated, in this case, by $8'$ to remove the background emission due to the atmosphere. The telescope array itself alternates between two ``Nod'' positions found symmetrically at a distance of $4'$ around the ``On'' source (or $8'$ from each other), thus creating two distinct ``Off'' positions for chopping \citep{Harper2018}. The beam size of HAWC+ at 214~$\mu$m is $18.2''$, or $5460$\,au at the distance of NGC~1333 at $300$~pc.

Polarization observations with HAWC+ are obtained from sets of four measurements taken with different half-wave plate positions (5.0$^{\circ}$, 50.0$^{\circ}$, 27.5$^{\circ}$, and 72.5$^{\circ}$; see \citealt{Harper2018}). These raw data files (Level~0) were downloaded from NASA's Infrared Science Archive (IRSA) and were then reduced and mosaicked into the final (Level~4) data product in July~2023 using the default parameters of the SOFIA Data Reduction software \citep[SOFIA Redux version 1.3.0;][]{melanie_clarke_2023}. This Level~4 data product has a pixel size of $4.55''$ and it contains the measured Stokes $I$, $Q$, and $U$ parameters, as well as derived quantities such as the de-biased polarization intensity $I_p$, the de-biased polarization fraction $P$, and the rotation angle $\theta_p$. The calculations for these quantities and their uncertainties are described in details by \citet{Gordon2018} and \cite{melanie_clarke_2023}.
 
\subsection{Dust Polarization at 850 $\mu$m with POL-2}\label{sec:JCMT}
We use published 850 $\mu$m dust polarization observations of NGC~1333 with POL-2 at the James Clerk Maxwell Telescope (JCMT) \citep{doi2020}, which were originally acquired for the B-fields In STar-forming Region Observations (BISTRO) survey \citep{Ward2017ApJ}. POL-2 is the polarimeter installed in front of the Submillimetre Common-User Bolometer Array Two (SCUBA-2), and it consists of a rotating half-wave plate and a polarizing grid \citep{Friberg2016SPIE}. SCUBA-2 is a 10,000 pixels bolometer array capable of simultaneous observing at 450 and 850~$\mu$m \citep{Holland2013MNRAS}. The 15-m dish of the JCMT provides SCUBA-2 with an effective resolution of $14.1''$ (i.e. 4200 au at the distance of NGC~1333 at $300$~pc) at 850~$\mu$m \citep{Dempsey2013MNRAS}.

These 850~$\mu$m polarization measurements of NGC~1333 with POL-2 were obtained in scan mode using a $11'$-wide $\text{POLCV}_{\text{DAISY}}$ pattern toward two partially overlapping pointings. The observations were acquired at a scanning speed of 8$\arcsec$ s$^{-1}$ and a data sampling rate of 200~Hz while the half-wave plate rotated at a frequency of 2~Hz. The data were reduced in November~2018 \citep[see][for details]{doi2020} using the software package STARLINK \citep{Currie2014ASPC}, and specifically the procedure \textit{pol2map} which is a variation on the script \textit{makemap} \citep{chapin2013}. The Stokes~$I$, $Q$, and $U$ parameters and their variances were initially produced on a grid of $4''$ pixels, and subsequently re-sampled to a $7''$ grid. These observations were used to make a quantitative comparison of B-fields with geometries obtained from VGT using $\rm N_{2}H^{+}$ line.

\subsection{Molecular line data} \label{sec:molecular_lines}
For the velocity gradient analysis, we use $^{12}$CO (3-2), $^{13}$CO (3-2), and C$^{18}$O (3-2) data from the James Clerk Maxwell Telescope (JCMT) science archive\footnote{Data can be accessed at https://www.cadc-ccda.hia-iha.nrc-cnrc.gc.ca/.}. NGC~1333 was observed in 2007 under program ID M06BGT02 (PI: Richard Hills) using the Heterodyne Array Receiver Program (HARP) instrument. 

We also use N$_2$H$^+$(1-0) observations from Combined Array for Research in Millimeter-wave Astronomy (CARMA) via the CARMA Large Area Star-formation SurveY (CLASSy; \citealt{2014ApJ...794..165S}). This survey used CARMA to simultaneously observe protostellar regions with both total power (single dish) and interferometric modes to achieve high resolution images that allows for the full recovery of large-scale emission. In this survey, 800 square arcminutes of the Perseus and Serpens molecular clouds were observed with CARMA, including the NGC\,1333 region studied in this work. These data were originally presented in \citet{Dhabal2018, Dhabal2019} and were shared with us by the CLASSy team via private communication. The resolution of the observations is 
$\rm 7\farcs2\,\times\,5\farcs4$ (2160\,au\,$\times$\,1620\,au at the distance of NGC1333).

\section{Method and Analysis}
\label{sec:analysis}

The main purpose of this work is to calibrate the Velocity Gradient Technique (VGT) using dust polarization measurements. First, we provide a synoptic review of the basis of the VGT and its development in light of recent theoretical findings. This synoptic review will motivate the application of the technique for this paper.

\subsection{Velocity Gradient Technique and the Gradient Statistics}

The VGT was first proposed by \citet{GL17} to study the statistical correlation between gradients of synthetic observables and B-field orientations. They numerically found that the gradients of column density and velocity centroid in synthetic observations from triply periodic isothermal compressible magnetohydrodynamic simulations are statistically perpendicular to those of the B-field. This statistical study presents similarities to the earlier work from \citet{soler2013}, where the latter found that the intensity gradients are statistically correlated to the B-field in the low column density limit. However, neither \citet{soler2013} nor \citet{GL17} could make predictions on B-field orientations since they are comparing the statistical properties between the gradients of a certain observable to the B-field only in a global sense.

\citet{GL17} suggested that the reason why both velocity centroid and column density gradients are correlated to the B-field orientation is due to the presence of Goldreich-Sridhar turbulence \citep{GS95}. In the case of strong turbulence (see \citealt{2023arXiv230106709Z} for the difference between weak and strong turbulence), Alfv\'enic turbulent eddies are anisotropic and scale dependent \citep{2002PhRvL..88x5001C}. Pictorially, the turbulent eddies are ellipsoids with their major axes aligned to the local direction of B-field \citep{2000ApJ...539..273C}. A later study \citep{CL03,2020PhRvX..10c1021M} found that the anisotropy of turbulent eddies are also present for slow magnetosonic modes, and, more importantly, Alfv\'en and slow modes occupy the majority of the turbulent energy in the interstellar medium \citep{2020NatAs...4.1001Z}. 

The locality property of magnetohydrodynamic turbulence opens up an opportunity to map B-field orientations in a smaller region, and potentially an orientation distribution rather than a global average. Indeed, \citet{YL17a} proposed a criterion to map B-field orientation via statistical properties of velocity centroid gradients: when the sampling area is ``sufficiently" large, the orientation histogram of velocity centroid gradients coincides with that of B-field directions. The peak of the velocity centroid gradients orientation histogram is therefore a measure of the B-field orientation in the sampling area, with an offset of $90^\circ$. The actual theoretical meaning of being ``sufficient" in statistics is later explored in \citet{LY18a}; in summary, the length scale that makes two orientation histograms similar corresponds to the turbulent correlation length scale.  Most importantly, \citet{YL17a} found numerically that the condition for ``sufficient" sampling is reached when the orientation histogram of velocity centroid gradients is Gaussian-like (see \S \ref{ssec:update} for issues related to the Gaussianity of gradient orientations). That means one could determine the minimal size of sampling by simply inspecting the shape of the gradient orientation histogram, opening a new way of mapping B-field directions without the need of polarization observations.

Subsequent publications after \citet{YL17a} follow two strategies of development. (i) The orientation histogram contains more information than just the peak of the distribution. For instance, the dispersion of the histogram gives the Alfv\'enic Mach number \citep{dispersion} and the amplitude of the histogram gives the sonic Mach number \citep{GA}. (ii) There are observables other than the gradients of velocity centroid or column density that better trace B-fields \citep{YL17b,LY18a,LY18b}. All these techniques are used in \cite{HuNatur2019}, where the B-field orientation of a background high velocity cloud {\it Smith Cloud} is predicted by the technique.

\subsection{Updates to the Gradient Theory}
\label{ssec:update}
\subsubsection{The correction from the Gaussianity condition \citep{YL17a}}
The most recent theoretical update on the VGT is based on the statistical properties of two-point correlation or structure functions (\citealt{2020MNRAS.496.2868L,ch9}, see also \citealt{LP00,LP04,LP12,LP16,KLP16,KLP17a,ch5})\footnote{We, however, acknowledge that the two-point statistical theory does not represent all of the MHD fluctuations in the sky. It is found in \cite{4DFFT,2023arXiv230106709Z} that a significant amount of MHD fluctuations are actually not the three MHD modes, but rather some nonlinear fluctuations called `non-waves'. \cite{4DFFT_theory} discussed the origin of this issue and gave the close form two-point theory that can be used under the framework of \cite{LP12}. However, how the emergence of `non-waves' changes our predictions, particularly the gradient technique, is still a subject of active research. See \cite{4DFFT_theory}.}, which attempted to explain the {\it empirical findings} that the gradient orientation histogram of a velocity-related observable is statistically Gaussian \citep{YL17a} {\it when the Alfv\'en Mach number $M_A$ is small}. \cite{2020MNRAS.496.2868L} pointed out that the gradient orientation distribution is not Gaussian when $M_A\sim 1$, but rather a special function predicted by the turbulence statistics theory. The use of Gaussians in the $M_A\sim 1$ case will lead to an underestimate of both Alfv\'enic Mach number\footnote{We do not suggest that the numerical finding from \cite{dispersion} and the applications thereafter are incorrect. From a theoretical point of view, the fitting from \cite{dispersion} could not be explained by the statistical theory of MHD turbulence, nor being further explained via the classical representations of MHD turbulence. Another theoretical approach, albeit being more constrained in terms of applications, is the curvature estimation from \cite{curvature}. When applying the techniques of velocity gradients, readers should clearly understand whether the technique is theoretically or empirically founded.} and polarization fraction. 

\cite{ch9} provided a first-principles theoretical analysis of how the directions of the B-field predicted by the gradient technique $\widetilde{\theta}_B$ and the magnetization $M_A = \sin\gamma\sqrt{-\log(J_2)/2}$ are given by the gradient orientation histogram. The theoretical distribution of gradient orientation {\kh for any 2D observables} is given by \citep{2020MNRAS.496.2868L}:
\begin{align}
\label{eq:Gaussian-grads}
P(\theta_p) &= \frac{1}{\pi} \times \frac{ 
\sqrt{1 - J_2}}
{1- \sqrt{J_2} \cos2 \left(\widetilde{\theta}_B-\theta_p\right)} .
\end{align}
This distribution function has two parameters: $J_2$ and $\widetilde{\theta}_B$. The first is the rotation-invariant ratio of the determinant of the traceless part of the covariance matrix and half of the trace of the covariance, which is
\begin{equation}
J_2 \equiv \frac{\left(\widetilde{\sigma}_{xx}-\widetilde{\sigma}_{yy}\right)^2 + 4 \widetilde{\sigma}_{xy}^2}
{\left(\widetilde{\sigma}_{xx}+\widetilde{\sigma}_{yy}\right)^2} ,
\label{eq:J2}
\end{equation}
where $\widetilde{\sigma}_{xx} = \langle \partial_x C \partial_x C\rangle$, $\widetilde{\sigma}_{xy} = \langle \partial_x C \partial_y C\rangle$, and $\widetilde{\sigma}_{yy} = \langle \partial_y C \partial_y C\rangle$ for any 2D observable C (C can be column density, velocity centroid, velocity channel, or velocity caustics,  i.e., velocity fluctuations arise from velocity channels by setting the 3D density to be constant, see \citealt{LP00} and \S \ref{sec:method}.). The prediction of the B-field (offset by 90$^\circ$ from gradient orientations, \citealt{GL17,YL17a}) and exceptional cases\footnote{\kh Gradients of 2D operators are not statistically necessarily perpendicular to the local B-field. For instance, gravity tends to turn the gradients of 2D observables to be parallel to the B-field, with density observables being more sensitive than velocity observables \citep{YL17b, hu2020}. Fast modes in certain orientations also exhibit similar behavior \citep{LY18a}.} by the gradient technique are given by \begin{equation}
\tan 2 \widetilde{\theta}_B \equiv \frac{2 \widetilde{\sigma}_{xy}}{\widetilde{\sigma}_{xx}-\widetilde{\sigma}_{yy}} .
\end{equation}

The applications of the statistical theory of MHD turbulence naturally prohibit the use of any ``Alignment Measure Improvement'' techniques (see, e.g., \citealt{LY18a}, \citealt{yue2019Nat}). The deviations of the gradients from the B-field are an intrinsic feature rather than noise or errors in the context of MHD theory\citep{ch5,ch9}. For example, the alignment measure itself, {\kh defined as $\langle \cos 2\Delta \phi\rangle$ where $\Delta \phi$ is the difference between the actual and the gradient-predicted B-field orientations \citep{GL17,YL17a}}, is actually a function of $M_A$ and predicted from theory \citep{LY18a}. Moreover, further smoothing or toggling of the statistics will cause the prediction of the B-field properties to significantly diverge from the theoretical predictions \citep{ch9}.

For the sake of this work, we will revert the gradient technique to the case where we can exactly describe it with the turbulence statistical theory {\it without the use of unsolicited parameters} (Filippova et al., in prep.). Moreover, \cite{ch9} also discussed that the gradient statistics can only be used in the \cite{YL17a} form for the estimation of the B-field strength.


\subsubsection{Retrieving Velocity Anisotropy from Velocity Caustics}
\label{sssec:caustics}
Caution needs to be taken on the choice of observables. The earlier development of the gradient technique focused on the statistical mean of both column density and velocity centroid gradients \citep{YL17a,YL17b}. However, from the perspective of MHD turbulence theory, density anisotropy can be very different from that of velocity, particularly in the case of low plasma $\beta$ media. For instance, density features are preferentially perpendicular to the B-field directions when the sonic Mach number is larger than 1. Velocity anisotropy in this regime is still preferentially parallel to the B-field \citep{2010ApJ...720..742K}. The separation of density and velocity fluctuations is therefore crucial for accurately predicting the direction of the B-field since column density provides only density fluctuations, while the velocity centroid provides both density and velocity fluctuations even after normalization \citep{KLP16}.

\citet{LP00} developed the theory relating the statistics of spectroscopic intensity fluctuations in position-position-velocity (PPV) space to statistical fluctuations of density and turbulent velocities. This theory describes the emergence of intensity fluctuations in PPV space due to velocity crowding, which are analytically studied by \cite{VDA}. Historically, \citet{LP04} termed the concept velocity caustics only in the absence of thermal broadening. This concept of velocity caustics in PPV was used by \citet{LP00} to denote the effect of velocity crowding due to multiple turbulent velocities along the line of sight. Due to this effect, the atoms that are at different physical positions along the same line of sight happen to overlap when viewed in PPV space. One of the most important consequences of the effect of velocity caustics is the creation of PPV intensity structures that are independent of the true density structures in 3D. As a result, there are non-zero PPV intensity fluctuations in spectroscopic data even from incompressible turbulence or in observational maps unrelated to velocity fluctuations. The ability to retrieve velocity caustics from observations would therefore provide a pathway to tracing velocity anisotropy in the sky, and thus better trace B-fields in interstellar media.

A careful investigation of velocity channel maps suggests that velocity fluctuations indeed exist, albeit they could be subdominant compared to density fluctuations in velocity channels \citep{VDA}. For our analysis on NGC~1333, we shall assume the turbulence is in an isothermal equilibrium state with the influence of gravity (see, e.g., \citealt{YL17b,LY18a}, \citealt{hu2020_selfgravity}) and outflows \citep{stephens2017}. In the current work, we follow the \cite{VDA} recipe and compute the gradients of pure velocity fluctuations.

\subsubsection{Differences in Line-of-Sight Summations Between Gradients and Polarization}

There is an additional disparity in how the gradient-predicted B-field vectors on each channel are added up along the line-of-sight. Intrinsically, Stokes parameters collect B-field fluctuations in a quadratic fashion ($(Q,U)\propto \int n_{\text{dust}}(\cos,\sin)(2\theta)$), while B-field direction from VGT are added up linearly ($(Q,U)_{VGT}\propto \int n_{\text{dust}}(\cos,\sin)(\theta)$)\citep{LY18a}. The addition of predicted B-field vectors by mimicking the addition of Stokes parameters is technically equivalent to a low-pass filter on the B-field statistics along the line-of-sight \citep[see][for a dedicated discussion]{ch9}. We will therefore restrain our gradient analysis to individual velocity channels.

\begin{figure}  
\includegraphics[width=0.49\textwidth]{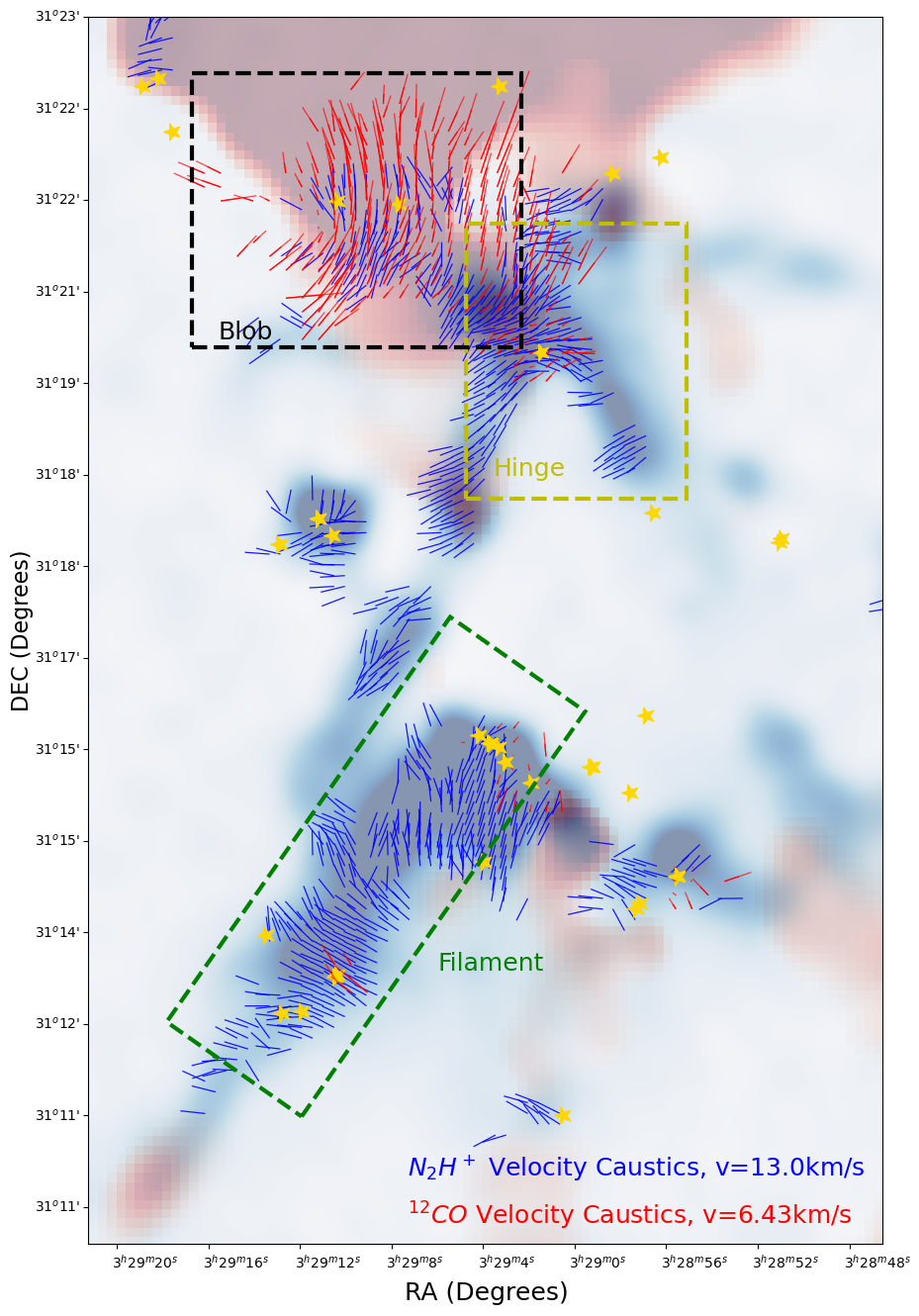}
\caption{The image shows the $N_{2}H^{+}$ and $^{12}CO$ velocity caustic maps in blue and red shades, respectively. Classification of regions studied in this paper are also identified and labelled.}
\label{fig:region}
\end{figure}


\subsection{Applying the VGT}\label{sec:method}
Here we describe the different techniques that we apply to the velocity cubes for the VGT analysis. \linebreak

{\noindent \bf Gradient Distributions and Block-Averaging:} Gradients were calculated following the procedures described in \cite{YL17a,YL17b} and improved by \cite{2020MNRAS.496.2868L},  including the gradient calculation method and the {\it sub-block averaging}, i.e., (i) divide the whole map into sub-blocks; (ii) compute the gradient orientation histogram within each sub-block, and (iii) estimate the B-field orientation within each sub-block. The {\it sub-block averaging} method is designed to define the most probable B-field direction within a block, which is a chosen 2D region with a definite size, by a local Gaussian/\citeauthor{2020MNRAS.496.2868L}-type fitting peak on the distribution of the gradient angle distribution. The use of sub-blocks increases the signal-to-noise of the gradients at the cost of spatial resolution of the spectral line observations. The uncertainties on the fit can provide a quantitative estimate on whether the block size is large enough to provide the probable direction. Here, we select a block size of $30\times 30$ pixels (on $\rm ^{12}CO$ map) for our analysis. This corresponds to $7.1'\times 7.1'$ or 0.6 pc$\times$ 0.6 pc at a distance of 300~pc of NGC1333. The estimated direction of the B-field is then obtained by adding $\pi/2$ (90$^\circ$) to each block-averaged channel gradient vector \citep{GL17,YL17a}.\linebreak

{\noindent \bf Velocity Channels and Centroids as Observables:} The \textcolor{red}{channel intensity} $p$, including the thermal broadening effect, is given by:
\begin{equation}
\begin{aligned}
p(\mathbf{X},v_0,\Delta v) = \int_{v_0-\Delta v/2}^{v_0+\Delta v/2} dv \,  \tilde{\rho}(\mathbf{X},v)\\
\times\left(\frac{m}{2\pi k_BT(\mathbf{X},v)}\right)^{1/2} e^{-\frac{mv^2}{2kT(\mathbf{X},v)}},
\label{eq:rhov_PPV}
\end{aligned}
\end{equation}
where the sky position is described by the 2D vector $\mathbf{X}=(x,y)$, the line-of-sight velocity $v_0$, and the channel width $\Delta v$ (See Tab.\ref{tab:symbols} for the definition for the rest of the symbols). The first moment of the integral of $p$ along the velocity axis, called the velocity centroid $C = \int dv \, v \, p({\bf X},v)$, is also found to be an observable that is more velocity dominant (i.e., velocity fluctuations dominate over density fluctuations in terms of the standard deviation, see \citealt{esquivel2005, esquivel2010, YL17a,VDA}).
\linebreak

{\noindent \bf The Velocity Decomposition Algorithm (VDA) for Subsonic Turbulence:}
As mentioned in \S \ref{sssec:caustics}, the gradient technique was originally designed for velocity-dependent observables, i.e., variables that are directly influenced by the velocity fluctuations along the line-of-sight. \cite{VDA} discussed that there is a way of reconstructing {\bf the velocity caustics $p_v$ } from each channel based on the linear algebra formalism. This formalism is known to work in subsonic media, and in supersonic media to an extent \citep{VDA}. To summarize, the {\bf density fluctuations $p_d$} and velocity caustics $p_v$ can be obtained from the following formulae using the full PPV cube $p({\bf R},v)$ and the column density map $I({\bf R})$, where $\langle p\rangle_{{\bf X}\in A}$ represents the velocity channel averaged over a certain spatial area $A$:
\begin{equation}
\begin{aligned}
    p_v &= p - \left( \langle pI\rangle-\langle p\rangle\langle I \rangle\right)\frac{I-\langle I\rangle}{\sigma_I^2}\\
    p_d &= p-p_v\\
    &=\left( \langle pI\rangle-\langle p\rangle\langle I \rangle\right)\frac{I-\langle I\rangle}{\sigma_I^2} .
\end{aligned}
\label{eq:VDA_ld2}
\end{equation}\linebreak

{\noindent \bf Magnetization from $J_2$:} The magnetization of velocity gradients \citep{ch9} is given by: 
\begin{equation}
M_A = \sin\gamma\sqrt{-\log(J_2)/2} ,
\label{eq:ch9_ma}
\end{equation}
where $J_2$ is a parameter similar to the circular variance obtained from fitting the histogram of gradient orientations via Eq.\ref{eq:Gaussian-grads}, while $\gamma$ is the independently measured angle between the B-field direction and the line-of-sight, e.g., via \cite{leakage,malik2023} (See Section~\ref{sec:malik}). This formula is exact for pure Alfv\'en fluctuations on velocity centroids, and it is justified to extend it to general MHD fluctuations and other velocity observables as given in \cite{ch9}.

\section{Results}
\label{sec:result}

\begin{figure}[ht]
\includegraphics[width=0.49\textwidth]{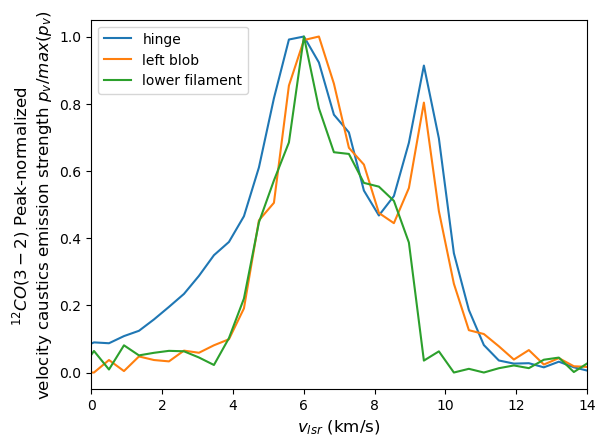}
\caption{Peak-normalized $^{12}$CO spectral line in the form of velocity caustics for the three separate regions. Hinge: blue; Blob: orange; Filament: green as shown in Figure~\ref{fig:region}.}
\label{fig:spectral}
\end{figure}

In this section, we present the results from applying the VGT described in \S \ref{sec:method} on NGC~1333. Based on all the available observations, there are three regions that show noticeably different B-field and velocity properties. As such, in our analysis we separate the map into three sub-regions: the ``Hinge", the ``Blob", and the ``Filament" (see Figure~\ref{fig:region}).

\subsection{Choice of Velocity Channel Positions}
For our analysis, we  first focus on the case of the $^{12}$CO spectral line, which is the spectral line detected over the largest spatial extent and which suffers more from absorption effects unless the density is very low (see, e.g., \citealt{2019ApJ...884..137H,2019ApJ...873...16H}). We note that absorption does not appear to be strong in NGC~1333 according to the channel maps of $^{12}$CO, $^{13}$CO, and C$^{18}$O. To proceed, we apply the VDA and plot the $^{12}$CO velocity caustics for the three separate regions. As seen in Figure~\ref{fig:spectral}, the three regions show clear difference in their velocity statistics. Notably, the ``Filament" has no velocity caustics fluctuations at $\sim$10\,km/s. Notice that the velocity caustics theory \cite{VDA} {\it predicts} the emergence of the ``double peak" spectral line, which is a unique feature for the velocity caustics fluctuations along the velocity axis. The velocity caustics tend to be crowded at the velocity position $|v-v_{peak}|\sim 1.06 \, \sigma_{turb}$, where $\sigma_{turb}$ is the turbulent velocity dispersion. The appearance of this double-peak feature is therefore a signature of a dominant turbulence system \citep{VDA}. This also indicates that the ``Filament" region may have some other mechanisms dominating the contribution due to turbulence, e.g., gravity or outflows \citep{stephens2017}.

\begin{figure*} 
\centering
\includegraphics[width=\textwidth]{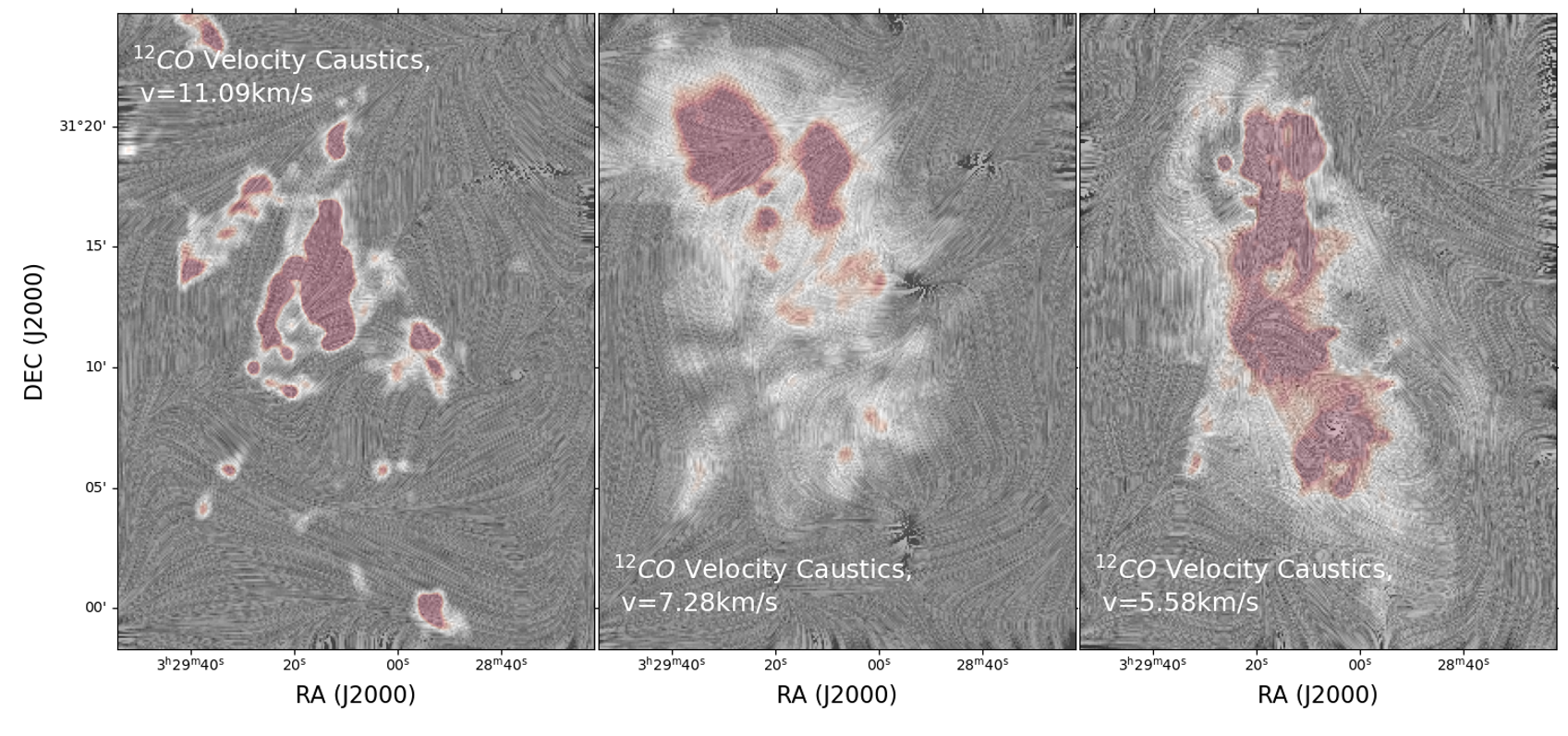}
\caption{The predicted B-field direction drawn via LIC given by the VGT method (\S \ref{sec:method}) overlaid on the velocity channel intensities for three selected velocity channels: $v=11.09, 7.28,  5.58$ km/s. }\label{Fig:polAv}
\end{figure*}

According to the recipe provided by \cite{VDA}, the velocity positions that correspond to the peaks of the velocity caustics of the spectral line (Figure~\ref{fig:spectral}) are preferred as the input for the gradient technique calculations. For our gradient analysis, we take these to be $5.58$ and $8.9$ km s$^{-1}$, which correspond to the double peaks of both the ``Hinge" and ``Blob" velocity caustics spectral lines, and $7.28$\,km\,s$^{-1}$, which corresponds to the spectral peak. Figure~\ref{Fig:polAv} shows the predicted B-field morphology (computed via velocity caustics) for these three selected velocity channels. Similar to Figure~\ref{fig:spectral}, we note that the channel maps display differences between velocities and species.

\begin{figure*}[ht]
\centering
\includegraphics[width=0.408\textwidth]{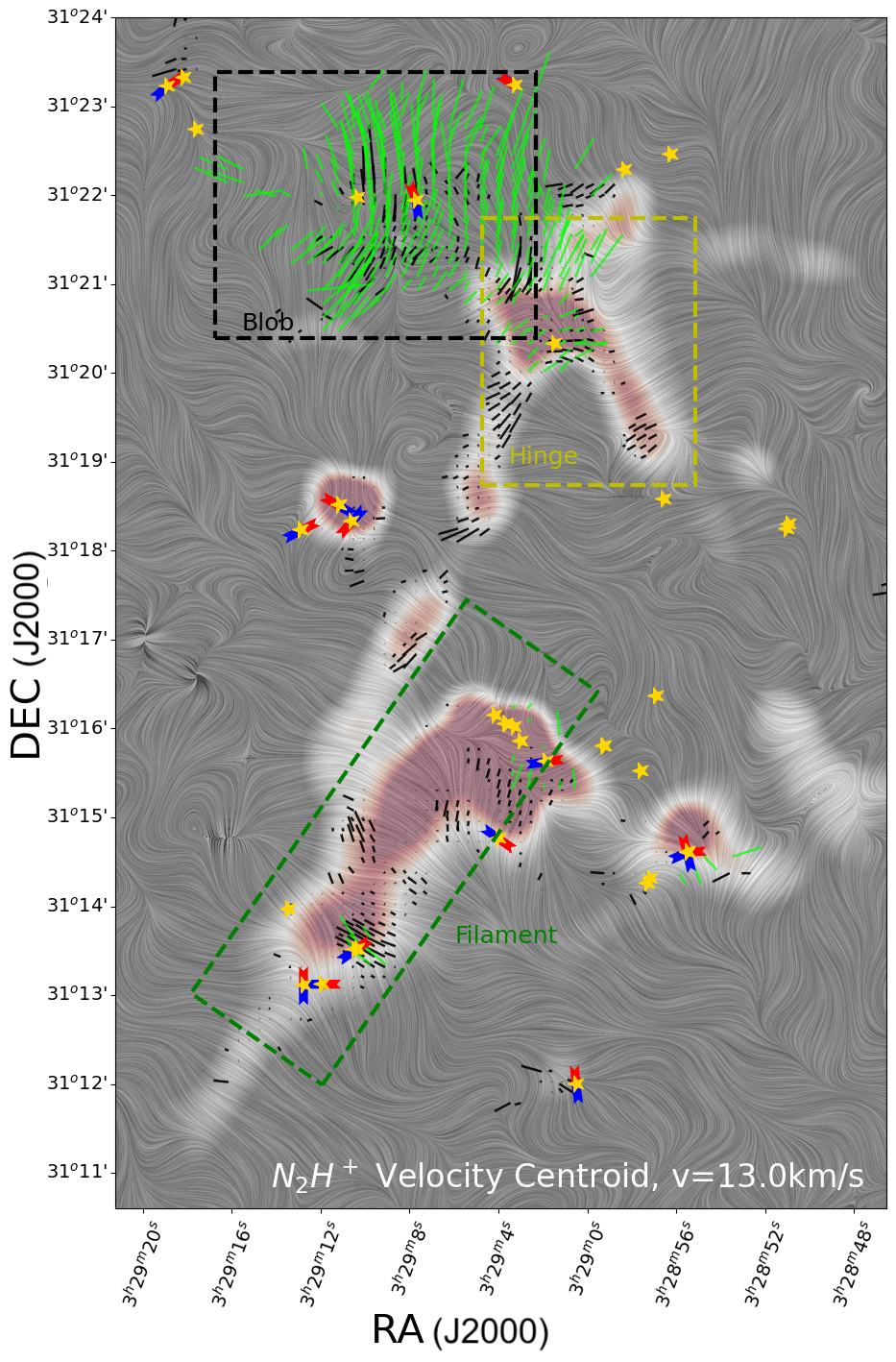}
\includegraphics[width=0.408\textwidth]{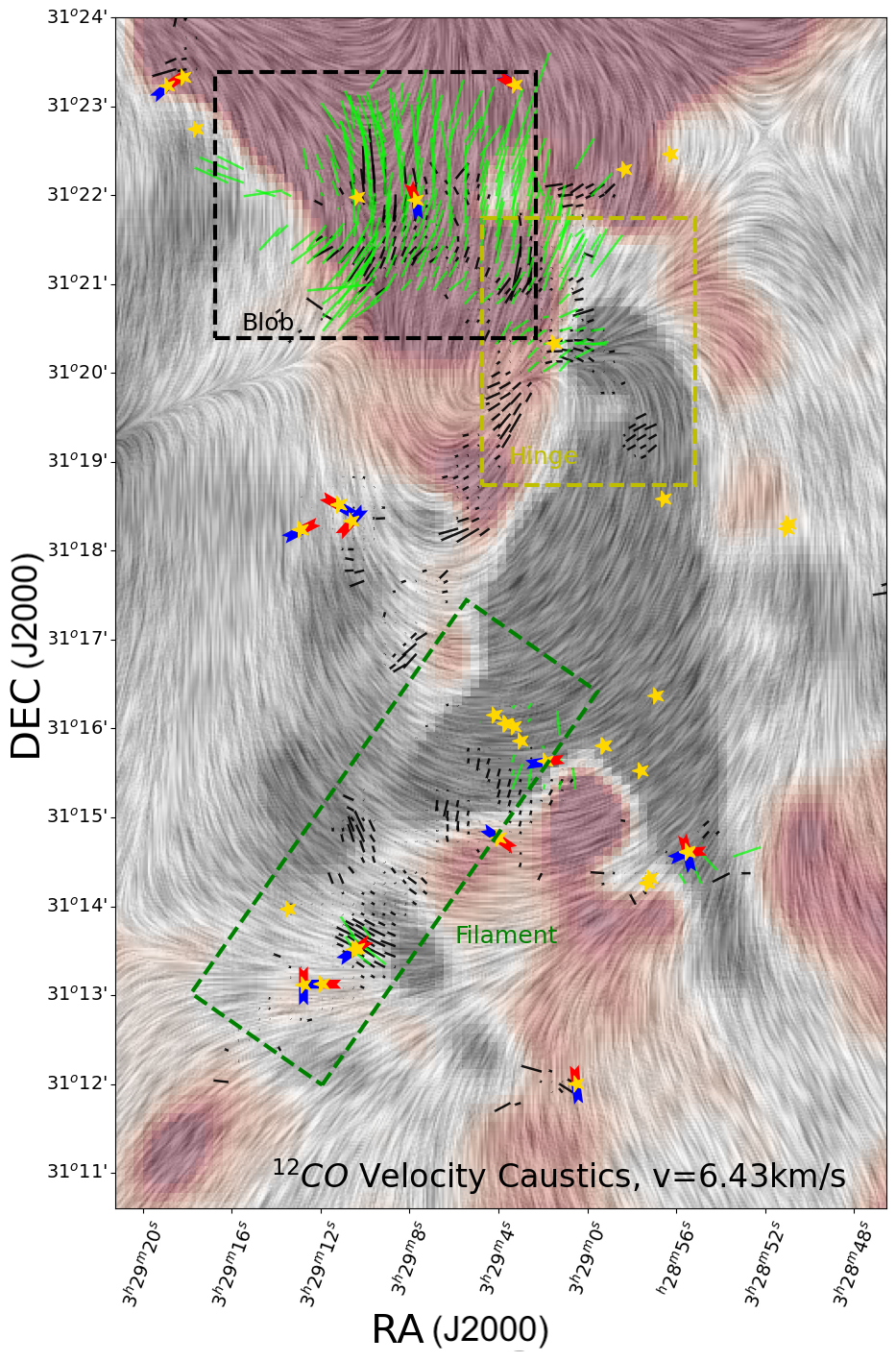}
\includegraphics[width=0.408\textwidth]{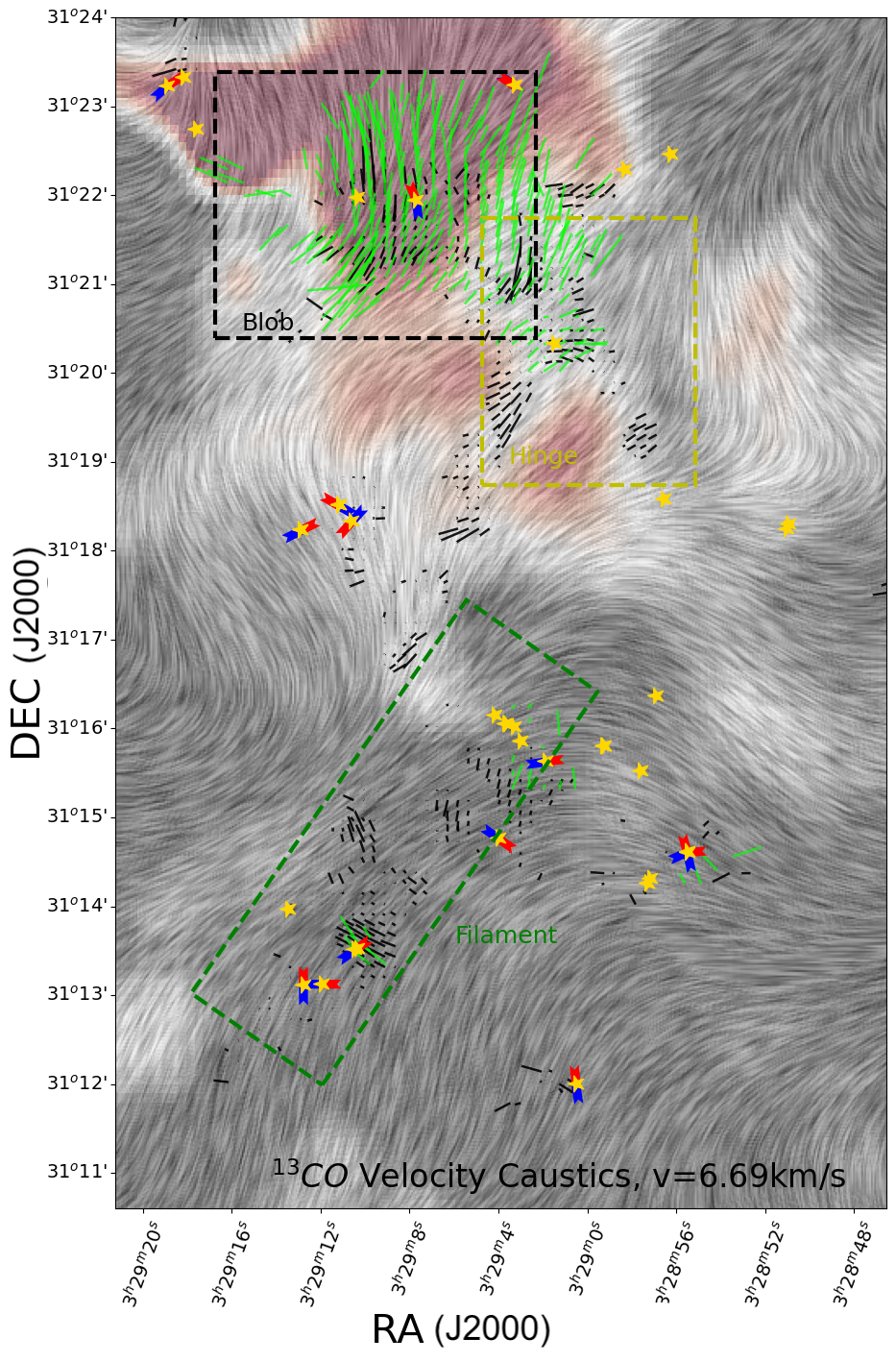}
\includegraphics[width=0.408\textwidth]{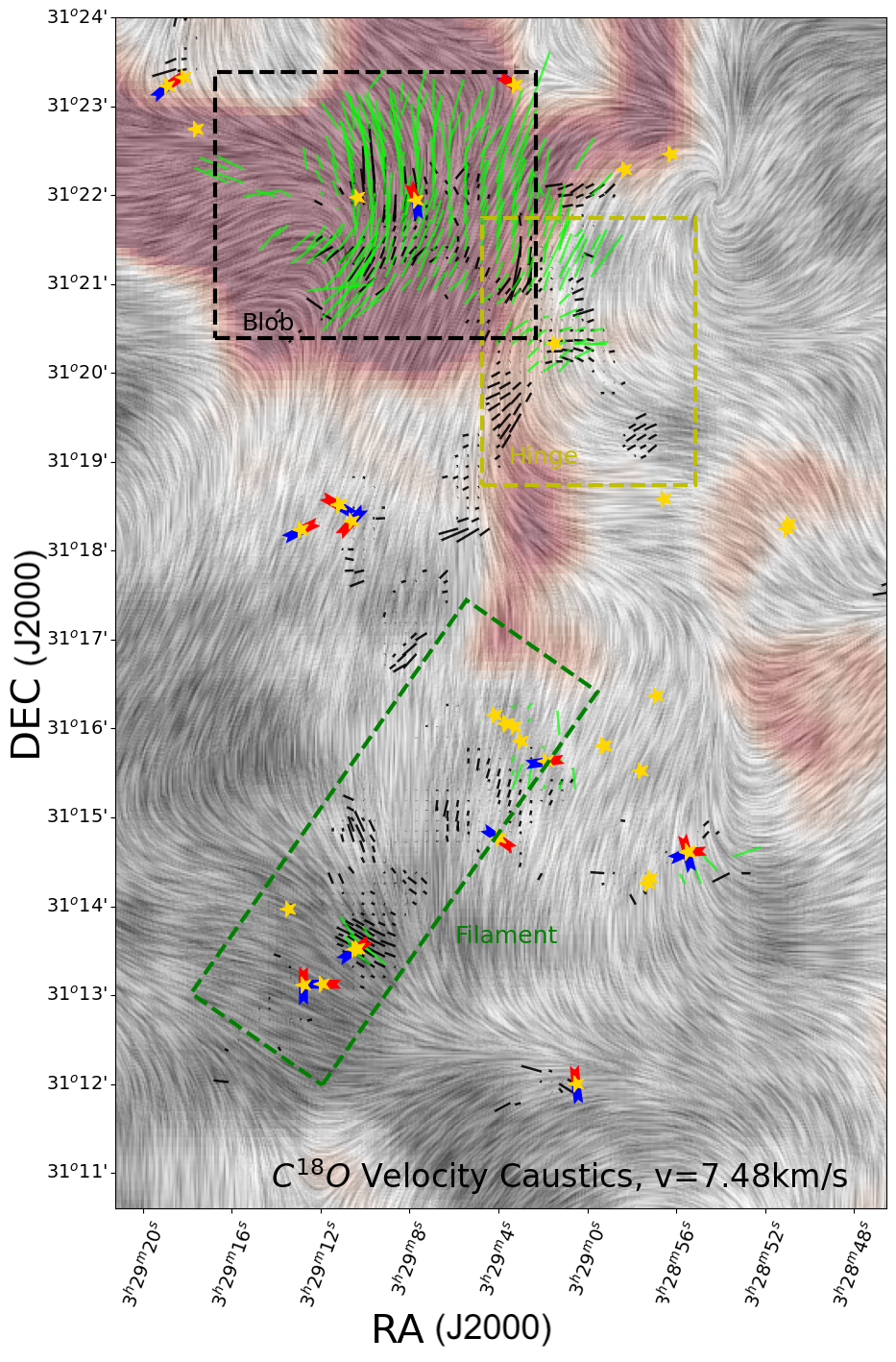}
\caption{(Top Left) The B-field direction computed at $6.43$\,km\,s$^{-1}$ as predicted by the N$_2$H$^+$ velocity gradient technique compared to the dust polarization at $214~\mu m$ (black vectors) and $850~\mu m$ polarization (green vectors). The yellow stars are the locations of the protostars \citep{tobin2016}. The region classifications (blob, hinge, filament) are laid out as in Fig~\ref{fig:region}. Furthermore, the outflow directions \citep{stephens2017} are plotted also in the same figure. Similar plots on $^{12}$CO, $^{13}$CO and C$^{18}$O channel maps are also provided at their respective velocity channel with peak intensities. The name of each tracer is labeled in the lower right corner.}
\label{Fig:caustics_216um}
\end{figure*}


\subsection{Plane-of-the-Sky B-field Orientation from the VGT}
\label{sec:B_lic}
We plot the B-field predicted by the VGT using data from both $^{12}$CO (Figures~\ref{Fig:polAv} and \ref{Fig:caustics_216um}) and N$_2$H$^+$ (Figure~\ref{Fig:caustics_216um}). To obtain these figures, we: (1) compute gradients from velocity caustics according to the recipe outlined in \S \ref{sec:method}, (2) produce the B-field map using the Line Integral Convolution (LIC,\citealt{10.1145/166117.166151}) implemented natively in Julia (see \citealt{LY18a}) with its parameter set to $200$ vectors, and (3) scale the output by the velocity channel intensity. Notably, Figure~\ref{Fig:polAv} shows that the predicted B-field directions can vary depending on the velocity channel used for the VGT. We then compare our gradient output with the polarization measurements at 214~$\mu$m and 850~$\mu$m, with the former restricted to the ``Blob" region only. Our analysis will mostly focus on the 850~$\mu$m polarization as it covers the three regions of interest listed previously.

Comparing the B-field morphology obtained from each velocity channel allows us to check whether the polarization vectors are tracing the same material and region as the gradients do. For example, in the top right panel of Figure~\ref{Fig:caustics_216um}, we plot the B-field directions predicted by the VGT {at $v=6.43$\,km\,s$^{-1}$} overlaid on the polarization vectors measured at 214~$\mu$m. This specific velocity channel was chosen for this panel as it shows a similar B-field morphology as the polarization data, and readers should note that the predicted B-field morphology using other channels, such as $5.58$\,km\,s$^{-1}$ or $7.26$\,km\,s$^{-1}$ as shown in Figure~\ref{Fig:polAv}, can show significant differences. Indeed, in the case of N$_2$H$^+$ in the top left panel of Figure~\ref{Fig:caustics_216um}, the 214~$\mu$m polarization does not match the directions predicted using the $v=13.0$\,km\,s$^{-1}$ channel map.

Here, we analyze the angle distribution obtained from the VGT and from the dust polarization. Since turbulence is a statistical measure, it must be compared with certain statistical parameters, including the mean and standard deviation. The standard method to compare the B-field orientation inferred from dust emission and from the VGT is via the alignment measure (i.e. AM), which spans from 0 to 1. In this case, the AM has a value close to 1, meaning there is good alignment between the overall B-field orientations inferred from the two techniques. However, this only provides a shallow conclusion and particularly masks many contributions from other mechanisms (e.g., gravity, outflow, shock) with significant information relating to the local star formation properties. Here, we remind the readers that the orientation of turbulent eddies concerning the local B-field is a statistical concept. Hence, in real space, the individual gradient vectors are not necessarily required to have any relation to the local B-field direction. Similar to sub-block averaging, one of the essential steps in applying the VGT is the expected value of the data within the block, which is thus the most probable orientation of the B-field direction \citep{YL17a}. 

We focus the analysis on three representative regions based on the integrated intensity map of the N$_{2}$H$^{+}$, as the tracer is tracing the gas more relevant to the cloud and star formation and without limitation from the depletion compare to C$^{18}$O or contamination of other structures as for $^{13}$CO and $^{12}$CO. We dubbed the three regions `blob', `hinge', and `filament'. The N$_{2}$H$^{+}$ emission of the ``Blob'' region is shown to have the least star formation activities and represents the region with the least dense gas. The `Hinge' region shows some star formation activity and stronger gas intensities. Finally, the `Filament' shows active star formation activities with hints of large-scale shocks (see Figure~\ref{fig:region}). 

\begin{figure*}[t]
\centering
\includegraphics[width=0.98\textwidth]{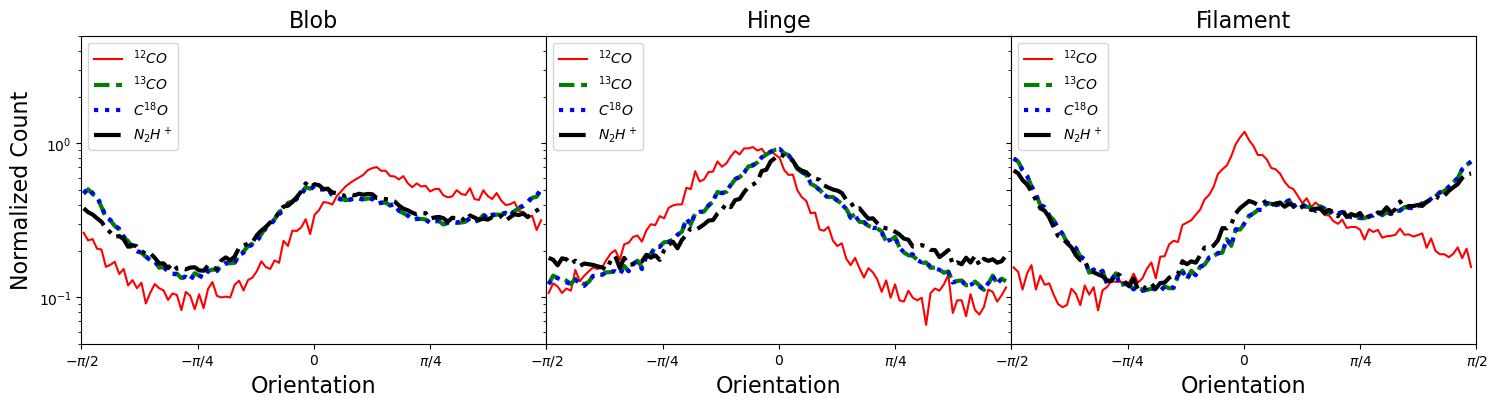}\\
\includegraphics[width=0.98\textwidth]{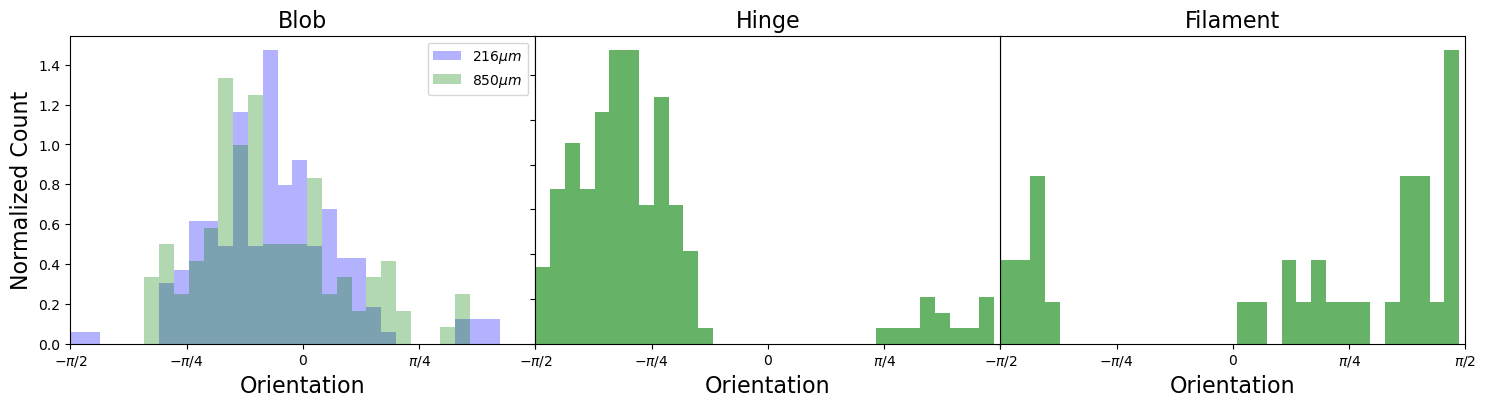}\\
\caption{\kh Histograms of both gradient angles (computed for each tracer in the upper panels) and polarization angles (computed for 214~$\mu$m and 850~$\mu$m dust polarization observations in the lower panels) for the three selected regions in NGC~1333, by choosing the same velocity channel. From left to right: Blob, Hinge, Filament. }
\label{fig:histogram}
\end{figure*}

We plot the histogram distribution of inferred B-field position angles for each region in the top panel of Figure~\ref{fig:histogram} for each tracer. When computing the histogram, we fix the velocity channel at the peak of N$_{2}$H$^{+}$. Notice that the gradient distribution from each molecular tracer can only trace the statistical properties of magnetized turbulence up to a certain optical depth \citep{2019ApJ...873...16H,2019ApJ...884..137H}. Here, we first compare the orientation inferred via VGT with that of different gas tracers. The shapes of the orientation distributions of different tracers are generally similar but with $^{12}$CO always offset from the other tracers from a few degrees in the `hinge' and `filament' regions to $\sim 20^{\circ}$ toward the `Blob' region. The rotated VGT angle histogram distributions of `Blob' and `Filament' regions have distinct peaks at $\pm 90^{\circ}$ and roughly at $0^{\circ}$, and a relative uniform distributions stretching from 0 to $90^{\circ}$ are observed. We also note that $^{12}$CO only present a single peak but not the other peaks at $\pm 90^{\circ}$. The `Hinge' region has a precise single peak shape at  $0^{\circ}$ but with a relatively large ($\sim 25^{\circ}$) dispersion. Compared to the magnetic field orientations inferred from polarized dust emissions, we generally find noteworthy features, and the main findings are summarised in the following.

We notice that, in general, there are no clear correlations between the orientations of rotated VGT and the dust-inferred magnetic field morphology based on polarized dust emissions except in the `filament' region. In particular, the `Hinge' region presents a bimodal distribution in dust polarization angles with peaks at $\pm 75^{\circ}$ (see lower panel of Figure~\ref{fig:histogram}), which is close to normal to the expected rotated VGT angles. This differences may be attributed to the effect of gravity, in which the rotated VGT orientations are normal to the dust polarization angle orientations. Turning to the `blob' region, the distribution between VGT and dust-inferred orientations is less distinct but still not consistent. While the dust polarization angle present a prominent peaks toward $\sim -25^{\circ}$, the rotated VGT angle histogram distribution shows more uniform distribution with a dip at $\sim 22.5^{\circ}$, suggesting clear misalignment with the dust polarization. Finally, we see more agreement between VGT orientation and the dust-inferred magnetic field orientations in the `filament' region. These results are in contrast to the prediction of anisotropy theory, which suggested an overall agreement between the rotated VGT orientations and the thermal dust inferred magnetic field orientations toward regions where gravity and feedback are not important, but with strong misalignment or even normal to the dust inferred magnetic field orientations when the effects of turbulent, feedback, and gravity become important. The `hinge' and `blob' are regions expected to have close alignment in the context of a dynamically important magnetic field regulating gas kinematics, but they instead display a clear misalignment larger than 20 degrees. On the other hand, the `filament' region, where the impact of feedback and gravitation are expected to be the greatest (e.g., \citealt{2022MNRAS.512.5214D}), ends up having the best alignment. The misalignment in the `hinge' and `blob' regions may be attributed to the highly turbulent environment, and thus its super-Alfv\` enic conditions. However, the magnetic field morphology is ordered with a light arced-shaped morphology. Therefore, it seems unlikely that the environment is super-Alfv\'enic, as the magnetic field morphology should be randomized. A possible explanation is that the curved magnetic field morphology could be results of nearby feedback, in which the compression alters the velocity caustic and thus resulting in the misalignment. However, there seems to be no clear feedback driving sources around the `Blob' region. For the `hinge' region, the large offset could be due to compression from nearby feedback (e.g., SVS13 region), resulting in the curved magnetic field morphology. The overall good alignment between the VGT directions (except ${}^{12}$CO) inferred from line measurements and the magnetic field direction inferred from dust toward the `filament' region suggest either the effects of feedback (e.g. shock, outflow, etc.) and gravity are not significant enough to alter the magnetic field morphology. However, the `filament' region has multiple identified outflows, shocks, and clear star forming activity \citep[see e.g.,][]{stephens2017b,2022MNRAS.512.5214D}, suggesting that the feedback should play a role in shaping the gas kinematics on the large scale that would be reflected in the VGT analysis. A possible explanation is that the shock propagation is parallel to the magnetic field orientation, thus the velocity gradient (affected by the shock) orientation align with the magnetic field \citep[see e.g. ][]{2022MNRAS.512.5214D}. The feedback in the filament may be more energetic than the B-fields, and thus the feedback may determine the direction of the B-fields observed via dust polarization. In summary, the applicability of VGT could be affected by shocks, and may be sensitive to the relative orientation between the shock propagation and the magnetic field direction, as explained below.  Toward the `hinge' region, the relative direction is close to 90 degrees between the magnetic field and the shock propagation. Since the velocity gradient is in the direction of the shock propagation, this results in a close to 90 degree shift between the VGT inferred direction and the magnetic field orientation. On the other hand, the relative orientation is parallel between the magnetic field direction and the shock propagation toward the `filament' region, which will result in a statistically good alignment between the VGT-inferred direction and the magnetic field direction, as reflected in the histogram distribution. However, in this case, the agreement just because the shock direction happen to be align with the magnetic field orientation. 

\begin{deluxetable}{c c c c }
\label{tab:result_ma}
\tablecaption{Statistical Measures of the gradient fluctuations. $M_A$ and $\gamma$ are defined by \cite{ch9,leakage,malik2023}. C stands for "Compressible mode dominated" while A stands for "Alfven mode dominated" case.} 
\tablehead{
& Blob  
& Hinge 
& Filament 
}
\startdata
$M_A/\sin\gamma$ & 0.74 & 0.52 & 0.69\\
$Y$ & 0.135 &  0.008 & 2.639\\
Inferred $\gamma$ & \makecell{$\gamma <5^o$ (C) \\or $\gamma>60^o$ (A)} &\makecell{$\gamma <5^o$ (C) \\or $\gamma>60^o$ (A)} &  $10^o < \gamma < 30^o$\\
Inferred $M_{A,3D}$ & \makecell{$<0.06$ (C)\\ $>0.64$ (A)} & \makecell{$<0.05$ (C)\\ $>0.45$ (A)}  & $0.12 - 0.34$
\enddata
\end{deluxetable}

\section{Inclination Angle $\gamma$, 3D $M_A$ and the 3-D B-field Properties of NGC~1333} \label{sec:malik}

The computation of the true 3D Alfv\'enic Mach number $M_A$ in the system requires additional information on how the mean B-field in the region of interest is inclined with respect to the line-of-sight. In a recent series of studies, \citet{leakage} and \citet{malik2023} provide a systematic method to retrieve the angle  $\gamma$ between the mean B-field and the line-of-sight from the statistics of the Stokes parameters. \cite{leakage} observed that the global correlation function of an observable $X$, denoted as $D_X$, is defined as follows:
\begin{equation}
D_X({\bf R}) = \int d^2 {\bf R}' (X({\bf R}') X({\bf R}'+{\bf R}))^2.
\label{eq:sf}
\end{equation}
where $R$ is the 2D vector on the plane-of-the-sky. Furthermore, \cite{malik2023} introduces a statistical technique known as the ``Y-parameter", which relies on the relative anisotropies of the correlation function of the observable quantities $I_{Stokes}+Q_{Stokes}\propto B_x^2$ and $I_{Stokes}-Q_{Stokes}\propto B_y^2$, where $B_x$ and $B_y$ represent the plane-of-sky components of the B-field. The Y-parameter is defined as:

\begin{equation}
Y = \frac{\text{Anisotropy}(D_{I_{Stokes}+Q_{Stokes}})}{\text{Anisotropy}(D_{I_{Stokes}-Q_{Stokes}})} = \frac{v/h(D_{I_{Stokes}+Q_{Stokes}})}{v/h(D_{I_{Stokes}-Q_{Stokes}})},
\label{eq_y}
\end{equation}

{where the terms $v$ and $h$ denote the minor and major axis of the correlation function relative to the projected magnetic field directions mapped by VGT}, respectively. The value of $\gamma$, together with the dominant fraction of B-field modes, can be calculated using the recipe outlined by \citeauthor{malik2023} (\citeyear{malik2023}, see \citealt{2023arXiv230713342M} for an application to the TeV Halo). We report our estimate of $\gamma$ in Table~\ref{tab:result_ma}. Readers can refer to \cite{malik2023} for a comprehensive diagnosis flowchart.  Together with the $M_A$ estimation method given in \S \ref{sec:method}, we provide our estimate of the 3D Alv\'enic Mach number $M_A$ in Table~\ref{tab:result_ma}. Our estimation of the 3D $M_A$ depends on whether the analyzed region is dominated by Alfv\'en or compressive turbulence, which we cannot accurately determine in this work. We therefore present in Table~\ref{tab:result_ma} the 3D $M_A$ and their respective $\gamma$ for both cases. Notice that the alfvenic mach is derived based on the ratio between the peak and the base \citep{dispersion}, which in this case the the VGT inferred orientations shows a prominent peak to apply for such estimation. 

\section{Discussion}
\label{sec:discussion}

\subsection{Caustics Gradients as primary B-field Tracers in Spectroscopic Observations}

\cite{VDA} resolves a number of misunderstandings in the community on what kinds of turbulence statistics channel maps contain. In general, a channel map can be written as a sum of two contributing parts: density and velocity. In particular, in the case of HI and other warm line tracers, the VDA can separate the contributions of the velocity and of the density when thermal broadening dominates over the turbulent contribution. 

Why is this separation of the density and velocity contributions so important, both theoretically and observationally? It is because the fundamental reason why gradients can trace B-fields is tied to their corresponding statistics in magnetized turbulence. According to the theory of MHD turbulence (e.g., \citealt{beresnyak_lazarian2006} and references therein), the velocity statistics are generally anisotropic along the B-field, while density statistics are often misaligned with the B-field when the underlying turbulent environment is in compression or self-gravitating. Indeed, it has repeatedly been shown by numerical papers \citep{YL17a, YL17b, LY18a, Hu2019c} that the gradients of velocity observables perform better than that of density gradients to trace B-fields. 

In contrast, \citet{LY18a} originally proposed that the velocity channels traces the fluctuations of pure velocity fluctuations. In that case, the velocity channel gradients should exhibit more similarity in terms of turbulence statistics to velocity fluctuations since the velocity channel gradients are better at tracing the B-field compared to the gradients of velocity centroids \citep{GL17, YL17a}. However, from a theoretical point of view, the pure velocity quantifier that traces B-fields the best are the velocity caustics, which is equivalent to the channel map in the limit of zero thermal broadening.


The VDA allows us to quantify the contributions of density and velocity fluctuations within velocity channels, and thus obtain velocity caustics that better represent reality. As demonstrated in \cite{VDA}, using caustics gradients we can focus on the pure velocity statistics in real applications without worrying about the contributions from densities where they have very different statistics. Together with the tool sets from \cite{VDA} ($1\sigma$ diagram, $P_d/P_v$ spectra), observers can now have full information on what scale and what velocity channels do they need to examine for a better accuracy of B-field tracing using the gradient technique.

\subsection{Caustics gradients under self-gravity}
A similar argument to the one presented in the previous subsection also applies when we are using gradients to probe self-gravitating regions. \citet{YL17b} suggested that gradients of both density and velocity features will {\it gradually}\footnote{The word ``gradually'' is important here since it is not a one-time process for the gradient vectors to flip their directions. Instead, as discussed in \cite{YL17b}, the relative orientation of both intensity and velocity centroid gradients will change from perpendicular to parallel according to the stage of collapse, with the latter being a slower process. } rotate from aligning perpendicularly to parallelly to the directions of the magnetic fields. With the VDA, we can further probe how the velocity caustics behave in the presence of gravity. According to the description in \cite{YL17b}, the caustic gradients should be the parameters least affected by gravity. However, when the gravitational force becomes dominant, the caustic gradients will eventually turn by 90$^\circ$. This change happens as the acceleration induced by gravity gets larger than the acceleration induced by turbulence. The relative orientations between the $p_v$ gradient and both the intensity gradient and the polarization vectors will change. This process is similar to the use of the velocity gradients (see \citealt{YL17b}), but we expect that using $p_v$ will help us describe the collapse regions significantly better. The corresponding methods and algorithms will be presented and discussed in a forthcoming paper (Yuen et al. in prep).

\subsection{Caustics gradients under the effects of outflows and shocks}
Outflows provide another source of velocity anisotropy that could affect the applicability of VGT. In particular, an outflow could drive and maintain turbulence, but it could also perturb the magnetic fields. NGC~1333 is particularly active with outflows \citep[e.g.,][]{stephens2017}. Here, it is unclear whether the VGT method is sensitive to the effects of outflows for the NGC~1333 region. However, $^{12}$CO better traces outflows than the other 3 spectral lines, which may in part explain why its orientation distribution (Figure~\ref{fig:histogram}) varies from the other spectral lines.

The good alignment between the VGT inferred orientation and the magnetic field morphology toward the active star-forming and feedback `filament' region raises the alarm on the validity of the application of VGT to infer magnetic field orientations. In fact, the comparison between caustic orientation and the magnetic field orientation reflects that the caustic gradient could be sensitive to the relative orientation between the shock propagation and the magnetic field direction. This also shows that large scale shocks arise from nearby HII region bubbles, supernova remnants (SNRs), or outflows (at a smaller scale) could have a significant impact on the caustic gradient. We stress that the applicability of VGT and its interpretation require a good characterization of the physical and chemical conditions of the region, as well as the relative importance of the contributions of the each mechanisms at different scales. A multi-scale view of the links between caustic gradient and magnetic field in connection to different physical mechanisms (e.g., shocks, outflow, gravity, etc.) toward NGC1333 will be presented and discussed in a forthcoming paper.


\section{Conclusion}
\label{sec:conclusion}
In this paper, we present the B-field morphology of NGC~1333 as traced by the Velocity Gradient Technique (VGT) and compared it with dust polarization measurements at 214~$\mu$m and 850~$\mu$m wavelengths. We divided the NGC~1333 cloud into three sub-regions according to B-field morphology and velocity gradients properties, namely the ``Blob", the ``Hinge", and the ``Filament" (c.f. Figure~\ref{fig:region}).  To summarize, we found that:

\begin{enumerate}
    \item The separation of intensity fluctuations within velocity channels is crucial for using the VGT. Particularly, the ``two-peak" phenomenon predicted by \cite{VDA} to be due to turbulent motions was observed in the ``Blob'' and ``Hinge'' regions (c.f. Figure~\ref{fig:spectral}). Differing from the other two regions, the velocity caustics distribution from the ``Filament'' region is not a two-peak distribution, meaning that turbulence might not be dominant in this sub-region of NGC~1333 \citep{VDA}.
    \item We apply the VGT on the {\it velocity caustics} maps of each selected sub-region for a range of chosen line-of-sight velocities $v$ (c.f. Figure~\ref{Fig:polAv}). 
    
    \item  We in general do not find a close correlation between the velocity gradient inferred orientations and the dust-inferred magnetic field orientations. The details are given below from the lowest best correlation:
    \begin{itemize}
        \item {\bf Hinge}: No clear correlations between the velocity gradient inferred orientations and the thermal dust inferred magnetic field orientations. The VGT inferred orientations show a clear single peaked distribution, while the thermal dust inferred magnetic field shows a close to normal bi-modal distribution. However we speculate the effect of shocks on the strong misalignment between caustic orientation and magnetic field direction.  

        \item {\bf Blob}: A less distinct angle distribution between the VGT and thermal dust inferred magnetic field angle, but the overall non-matching orientation distribution still unveils inconsistency between the angles inferred between the gas and dust.
        
        \item {\bf Filament}: The B-field structure traced with the VGT shows the best agreement, but this correlation could be due to the effect of shock.
        
        All in all, these findings are in contrast to the predictions from the typical VGT studies in the literature, thus requiring caution and revision to the assumptions and the results presented in the literature.  
    \end{itemize}  
    \item We estimate both the inclination angle and the $M_A$ of the three regions (See Tab.\ref{tab:result_ma}) and finds that these regions are all sub-alfvenic. 
\end{enumerate}


{\noindent\bf Acknowledgments.} The research presented in this paper was supported by the Laboratory Directed Research and Development program of Los Alamos National Laboratory under project number(s) 20220700PRD1, 20220107DR. We acknowledge Yasuo Doi and the BISTRO survey team, as well as the CLASSy team as a whole, for providing us with polarimetric and spectroscopic data for NGC~1333.

Based on observations made with the NASA/DLR Stratospheric Observatory for Infrared Astronomy (SOFIA). SOFIA is jointly operated by the Universities Space Research Association, Inc. (USRA), under NASA contract NNA17BF53C , and the Deutsches SOFIA Institut (DSI) under DLR contract 50 OK 2002 to the University of Stuttgart. Financial support for IWS and SC was provided by NASA through awards 06\_0098 and 08\_0186 issued by USRA. 

The James Clerk Maxwell Telescope is operated by the East Asian Observatory on behalf of The National Astronomical Observatory of Japan; Academia Sinica Institute of Astronomy and Astrophysics; the Korea Astronomy and Space Science Institute; the National Astronomical Research Institute of Thailand; Center for Astronomical Mega-Science (as well as the National Key R\&D Program of China with No. 2017YFA0402700). Additional funding support is provided by the Science and Technology Facilities Council of the United Kingdom and participating universities and organizations in the United Kingdom and Canada. Additional funds for the construction of SCUBA-2 were provided by the Canada Foundation for Innovation. The authors wish to recognize and acknowledge the very significant cultural role and reverence that the summit of Maunakea has always had within the indigenous Hawaiian community.

\vspace{5mm}

\software{LIC, MatPlotLib \citep{2007Matplotlib}, NumPy \citep{2011NumPy,2020NumPy}, AstroPy \citep{2013AstroPy,2018Astropy}}

\bibliographystyle{aasjournal}
\bibliography{ref}

\appendix

\begin{table*}[h]
\caption{Table of symbols used in this paper\label{tab:symbols}}
\centering
\begin{tabular}{cc}
\hline
\hline
Symbol & Meaning  \\
\hline\hline 
\multicolumn{2}{l}{\bf Shorthand Operators}\\
$\langle ...\rangle_{a}$ & Spatial averaging operator over parameter a.\\ \hline
\multicolumn{2}{l}{\bf Plasma Parameters}\\
$\rho$ & Fluid density.\\
${\bf v}$ & Fluid velocity vector, RMS value of it =$v_{turb}$.\\
${\bf B}_0$ & Mean B-field vector, its unit vector is sometimes denoted as $\hat{\lambda}$.\\
${\bf \delta B}$ & $={\bf B}-{\bf B}_0$, the mean-subtracted B-field.\\
$v_A$ & $=\frac{|{\bf B_0}|}{\sqrt{4\pi\langle \rho\rangle}}$, Alfvenic Speed\\
$c_S$ & Sonic Speed \\
$M_s$ & $= v_{turb}/c_s$, Sonic Mach number.\\
$M_A$ & $= v_{turb}/v_A$, Alfvenic Mach number.\\
$\beta$ & $=2M_A^2/M_s^2$, plasma compressibility.\\
$\gamma$ & The angle between the mean B-field and line of sight\\
${\bf X}$ & Plane of sky displacement vector\\ 
$\Delta v$ & Velocity Channel Width \\ 
 $T$ & Temperature\\
\hline
\multicolumn{2}{l}{\bf Statistical Parameters}\\
$p({\bf X},v_0,\Delta v)$ & Velocity channel at velocity $v_0$, with velocity channel width $\delta v$ \citep{LP00}.\\
$Y$ & Y-parameter, defined by Eq.\ref{eq_y} \citep{malik2023}.\\
$J_2$ & Gradient Angle Variance measured by the \citeauthor{2020MNRAS.496.2868L}-type distribution function, defined by Eq.\ref{eq:J2}.\\
$\theta_B$ & Predicted B-field angle by the \citeauthor{2020MNRAS.496.2868L}-type distribution function.\\
$D_X$ & 2D correlation function function for observable $X$.\\
\hline\hline
\end{tabular}
\end{table*}


\end{document}